\documentstyle[12pt]{article}

\begin{document}

\baselineskip = 20pt

\title{\bf Proper Time Method for Fermions}
\bigskip

\author{Ashok Das\\
Department of Physics and Astronomy \\
University of Rochester\\
Rochester, NY 14627, USA\\
\\
C. Farina$^{\dagger}$\\
Instituto de F\'\i sica\\
Universidade Federal do Rio de Janeiro\\
Rio de Janeiro, RJ 21945-970, BRASIL\\
\\}
\bigskip

\maketitle

\begin{abstract}

We show how Schwinger's proper time method can be used to evaluate directly 
the determinant of first order operators associated with fermionic theories. 
Several examples are worked out in detail.

\end{abstract}

\vfill

\noindent $^\dagger${e-mail: farina@if.ufrj.br}

\pagebreak

\section{Introduction}

\par

Schwinger's proper time method has been successfully used, in the past, 
to evaluate various 
one loop effects in quantum field theories \cite{Schwin51,Schwin92a,Lowe}
(see also the books by Greiner and Reinhardt \cite{Greiner} and by Reuter
and Dittrich \cite{Reuter} and references therein for more details; an
elementary introduction on Schwinger's proper time method can be found in
ref. \cite{Farina} ). It is a very simple and powerful 
method where the operator whose determinant is being evaluated is treated as
the Hamiltonian for evolution in an extra time direction known as the 
\lq\lq proper
time". One can easily evaluate the determinant by solving the \lq\lq Heisenberg
equations" for this quantum mechanical problem. However, when applied
directly to the evaluation of determinants of first order operators associated
with fermionic theories, this method 
runs into difficulties in the sense that the \lq\lq Heisenberg equations" 
do not lead to
convenient solutions which can be used to evaluate the determinant (We will
explain this in more detail in the following sections). In even space-time 
dimensions, one usually remedies this problem by 
converting the determinant of the first order operator to that of a second 
order operator through the use 
of the $\gamma_5$ matrix. However, in odd space-time dimensions, there is no 
$\gamma_5$ matrix and, therefore, the question of a direct evaluation of the 
determinant of first order operators becomes quite important.

In this paper, we show 
systematically, how one can use the proper time method to evaluate directly 
the determinants of first order operators associated with fermionic
theories. In section {\bf 2}, we evaluate 
the partition function for the bosonic oscillator which involves evaluating 
the determinant of a second order operator. We choose this simple example 
mainly to bring out the reasoning that goes into such an evaluation. In 
section {\bf 3}, we show how this method fails when applied directly to the 
evaluation of the partition function for a free fermionic oscillator which 
involves a first order operator. We also show, in this simple example, 
how it can be converted to a problem involving a second order operator and 
hence can be solved. In section {\bf 4}, we evaluate the effective action for 
a fermion interacting with an arbitrary, external electromagnetic field, 
in $0+1$
dimension, within the context of the proper time method. In section {\bf 5}, 
we further apply this method to solve the Schwinger model as well as 
the general Abelian model. Section {\bf 6} contains a brief conclusion.

\section{Bosonic oscillator}

Let us consider the bosonic 
oscillator in $0+1$ dimensions. The partition function for this simple system 
can be calculated in many different ways. However, we will use here the proper
time method for the evaluation of the partition function mainly to establish 
the steps that go into such an evaluation. The partition function is easily 
evaluated in the imaginary time formalism \cite{Dasbook}. The Euclidean
Lagrangian for the 
oscillator is given by (up to a total derivative)
\begin{equation}
L_E={1\over2}x\left(-{d^2\over dt^2}+m^2\right) x\; .
\end{equation}
Integrating out the variable $x$, in the path integral, leads to the partition 
function for the bosonic oscillator as 
\begin{equation}
Z_B(\beta)=\left[\det\left( -{d^2\over dt^2}+m^2\right)\right]^{-1/2}=
\exp\biggl\{ -{1\over2} {\rm Tr}\ln\left( -{d^2\over dt^2}+m^2\right)
\biggr\}\; .
\end{equation}
Here the determinant has to be evaluated in a space of functions periodic with 
an interval $\beta$ which is the inverse of the temperature (in units of the 
Boltzmann constant, $k_B=1$). Here \lq\lq Tr" stands for the trace to be taken 
over a complete basis.

In the proper time formalism, one represents (up to a
 constant independent of dynamics) for any (non-negative) operator, $H$, 
\begin{equation}
{\rm Tr}\ln H=-\int_0^\infty {d\tau\over\tau} {\rm Tr}\; e^{-\tau H}\; .
\end{equation}
Here $\tau$ is referred to as the proper time and the evaluation of the trace 
proceeds as follows. For the  present case, we note that we can identify ($p=-i
{d\over dt}$)
\begin{equation}
H=-{d^2\over dt^2}+m^2=: p^2+m^2\; ,
\end{equation}
and with $\hbar=1$,
\begin{equation}
[t,p]=i\; .
\end{equation}
Consequently, we can write
\begin{equation}
{\rm Tr} \; e^{-\tau H}=\int dt\; \langle t\vert e^{-\tau H}\vert t\rangle\; .
\end{equation}

The trace is evaluated by introducing the (modified) Heisenberg picture
\begin{eqnarray}\label{Heisenberg}
\vert t,\tau\rangle&=&e^{\tau H}\vert t\rangle\nonumber\\
\langle t,\tau\vert &=& \langle t\vert e^{-\tau H}\; ,
\end{eqnarray}
so that 
\begin{equation}\label{equacao}
\partial_\tau\langle t,\tau\vert t',0\rangle=-\langle t,\tau\vert 
H\vert t',0\rangle\; .
\end{equation}
Furthermore, the (modified) Heisenberg equations give 
\begin{eqnarray}
{dt\over d\tau}&=&-[t,H]=-[t,p^2+m^2]=-2ip\nonumber\\
{dp\over d\tau}&=&-[p,H]=-[p,p^2+m^2]=0\; ,
\end{eqnarray}
which can be solved to give
\begin{equation}\label{ptau}
p(\tau)=p(0)=-{t(\tau)-t(0)\over 2i\tau}\; .
\end{equation}
This also leads to the commutation relation
\begin{equation}\label{ttau}
[t(\tau),t(0)]=-2\tau\; .
\end{equation}

The relations (\ref{ptau}) and (\ref{ttau}) can be used to write
\begin{eqnarray}
H&=&-{\left( t(\tau)-t(0)\right)^2\over 4\tau^2}+m^2\nonumber\\
&=&-{1\over 4\tau^2}\left[ \left( t(\tau)\right)^2 -2t(\tau)t(0)+ \left( t(0)
\right)^2\right]+\left( m^2+{1\over 2\tau}\right)\; ,
\end{eqnarray}
so that we obtain
\begin{eqnarray}
\partial_\tau\langle t,\tau\vert t',0\rangle&=&-\langle t,\tau\vert 
H\vert t',0\rangle\nonumber\\
&=&\biggl\{ {1\over 4\tau^2}( t-t')^2 -\left( m^2+{1\over 2\tau}\right)\biggr\}
\langle t,\tau\vert t',0\rangle\; ,
\end{eqnarray}
which can be integrated to give
\begin{equation}
\langle t,\tau\vert t',0\rangle={C\over \sqrt{\tau}}\; e^{-\tau m^2-(t-t')^2/4
\tau}\;\; .
\end{equation}
Here $C$ is a constant of integration which can be determined from various 
consistency conditions as well as the requirement
\begin{equation}\label{delta}
\lim_{\tau\rightarrow 0}\langle t,\tau\vert t',0\rangle=\delta(t-t')\; ,
\end{equation}
so that we can write
\begin{equation}\label{Kzero}
\langle t,\tau\vert t',0\rangle=:K^{(0)}(t-t';\tau)=
{1\over \sqrt{4\pi\tau}}\; e^{-\tau m^2-(t-t')^2/4\tau}\;\; .
\end{equation}

So far, we have not worried about the periodicity condition necessary for the 
evaluation of the determinant. Let us next introduce
\begin{equation}\label{K}
K(t-t';\tau)=\sum_{n=-\infty}^\infty K^{(0)}(t-t'-n\beta;\tau)\; ,
\end{equation}
which can be easily seen to satisfy the required periodicity condition, namely:
\begin{equation}
K(t-t'+\beta;\tau)=K(t-t';\tau)\; .
\end{equation}
Thus, we can now write
\begin{eqnarray}\label{trace}
{\rm Tr}\ln\left(-{d^2\over dt^2}+m^2\right)&=&-\int_0^\infty
{d\tau\over\tau}\int_0^\beta dt\; K(0;\tau)\nonumber\\
&=&{\beta\over \sqrt{ 4\pi}}\sum_{n=-\infty}^\infty\int_0^\infty\; d\tau\,
\tau^{-3/2}\, e^{-m^2\tau-n^2\beta^2/4\tau}\; ,
\end{eqnarray}
where we used equations (\ref{Kzero}) and (\ref{K}). Identifying in the
previous equation an integral representation of the modified Bessel function
of second kind $K_{-1/2}(n\beta\omega)$ and using the fact that
$K_{-1/2}(z)=\sqrt{\pi/2z}\; e^{-z}$, the trace in (\ref{trace}) can be
evaluated in a straightforward manner to give 
\begin{eqnarray}
Z_B(\beta)&=&\exp\biggl\{-{1\over2}{\rm Tr}\ln\left(-{d^2\over dt^2}+m^2\right)
\biggr\}\nonumber\\
&=&\exp\biggl\{-{1\over2}\; 2\ln\left[ 2\sinh(\beta m/2)\right]\biggr\} 
\nonumber\\
&=&{1\over 2\sinh(\beta m/2)}\; .
\end{eqnarray}

This is indeed the correct partition function for the bosonic oscillator. We 
emphasize here that, in the proper time formalism, the crucial point lies in 
the fact that the Heisenberg equations can be solved to express the momentum 
in terms of the conjugate operators (see eq. (\ref{ptau})). As we will see
in the next section, 
this is not possible for operators which are first order in the momentum.

\section{ Fermionic Oscillator}

Let us next evaluate the partition function for the fermionic oscillator in the
proper time formalism. Once again, this can be evaluated in many different 
ways, but we choose this simple example to bring out the difficulties 
associated with
first order operators in the proper time formalism. Let us start with the 
Euclidean Lagrangian for the fermionic oscillator,
\begin{equation}
L_E=\overline\psi({d\over dt}+m)\psi\; .
\end{equation}
We can integrate out the fermions, in the path integral, to obtain the 
partition function for the fermionic oscillator as
\begin{equation}
Z_F(\beta)=\det\left({d\over dt}+m\right)=\exp\biggl\{ {\rm Tr}\ln
\left({d\over dt}+m\right)\biggr\}\; .
\end{equation}
Here the determinant is understood to be evaluated in the space of functions 
anti-periodic with an interval $\beta$.

Following the discussion of the 
previous section, we note that we have to solve, in this case,
\begin{equation}\label{eqdif}
\partial_\tau\langle t,\tau\vert t',0\rangle=-\langle t,\tau\vert 
H\vert t',0\rangle\; ,
\end{equation}
with
\begin{equation}
H={d\over dt}+m=ip+m\; .
\end{equation}
The (modified) Heisenberg equations, in this case, give
\begin{eqnarray}
{dt\over d\tau}&=&-[t,H]=-[t,ip+m]=1\nonumber\\
{dp\over d\tau}&=&-[p,H]=-[p,ip+m]=0\; .
\end{eqnarray}
These equations can be solved to give
\begin{eqnarray}
p(\tau)&=&p(0)\nonumber\\
t(\tau)&=&t(0)+\tau\; .
\end{eqnarray}
However, they do not allow us to express the momentum in terms of the conjugate
operator. Consequently, the evaluation of the matrix element on the right hand
side of (\ref{eqdif}) becomes extremely difficult. This is the difficulty with 
the first order operators in the proper time formalism.

In this simple example, we can relate the first order operator to a second
order
operator in a simple manner and thereby evaluate the partition function easily.
Let us note that (up to a total divergence)
\begin{eqnarray} 
L_E&=&\overline\psi({d\over dt}+m)\psi\nonumber\\
&=&{1\over2}\overline\psi({d\over dt}+m)\psi +
{1\over2}((-{d\over dt}+m)\overline\psi)\psi\; .
\end{eqnarray}
Thus, introducing
\begin{equation}
\Psi=\left(\begin{array}{c}\psi\\\overline\psi\end{array}\right)
\end{equation}
we can write
\begin{equation}
L_E={1\over2}\Psi^T\left(\sigma_1{d\over dt}-i\sigma_2 m\right)\Psi\; ,
\end{equation}
where the $\sigma$'s are the usual Pauli matrices. Integrating out the 
fermions, we obtain
\begin{eqnarray}
Z_F(\beta)&=&\left[\det\left(\sigma_1{d\over dt}-i\sigma_2 m\right)
\right]^{1/2}\nonumber\\
&=&\left[\det\left(\sigma_1{d\over dt}-i\sigma_2 m\right)^2\right]^{1/4}
\nonumber\\
&=&\left[\det\left({d^2\over dt^2}-m^2\right)I\right]^{1/4}\nonumber\\
&=&\left[\det\left(-{d^2\over dt^2}+m^2\right)\right]^{1/2}\nonumber\\
&=&\exp\left\{{1\over2}{\rm Tr}\ln\left(-{d^2\over dt^2}+m^2\right)
\right\}\; .
\end{eqnarray}

In this way, we have related the determinant (partition function) of the first 
order operator describing the fermionic oscillator to that of the second order
operator of the last section. The only difference is that the determinant, 
here, has 
to be evaluated over anti-periodic functions. Thus, we define
\begin{equation}
\tilde K(t-t';\tau)=\sum_{n=-\infty}^\infty (-1)^n K^{(0)}(t-t'-n\beta;\tau)
\; ,
\end{equation}
where $K^{(0)}(t-t';\tau)$ is defined in (\ref{Kzero}). It is easy to 
check that $\tilde K(t-t';\tau)$ defined above satisfies the required 
anti-periodic condition, namely,
\begin{equation}
\tilde K(t-t'+\beta;\tau)=-\tilde K(t-t';\tau)\; .
\end{equation}
Therefore, we obtain
\begin{equation}
{\rm Tr}\ln\left(-{d^2\over dt^2}+m^2\right)=-\int_0^\infty {d\tau\over \tau}
\int_0^\beta dt\; K(0;\tau)\; ,
\end{equation}
which can be evaluated in a simple manner to give 
\begin{eqnarray}
Z_F(\beta)&=&\exp\biggl\{{1\over2}{\rm Tr}\ln
\left(-{d^2\over dt^2}+m^2\right)
\biggr\}\nonumber\\
&=&\exp\biggl\{{1\over2} \; 2{\rm Tr}\ln\left[ 2\cosh(\beta m/2)\right]
\biggr\}\nonumber\\
&=&2\cosh(\beta m/2)\; .
\end{eqnarray}
This is indeed the correct partition function for the fermionic oscillator.

\section{Fermion in an External Field}

In the last section, we evaluated the partition function for a free fermionic 
oscillator 
in the proper time formalism by relating the determinant of a first order 
operator to that of a second order operator for which the Heisenberg equations
 can be solved in a required form. However, this cannot always be done and so 
a direct evaluation of the solution of eq. (\ref{equacao}) becomes essential. 
In this section, we analyze the case of a fermion interacting with an
arbitrary, external
 gauge field in $0+1$ dimension. The Lagrangian is given by (Minkowski space)
\begin{equation}
L=\overline\psi(i\partial_t-m-A)\psi\; ,
\end{equation}
where $A$ is an arbitrary function of $t$ and we are interested in evaluating 
the contribution of the fermions to the effective action which is given by
\bigskip
\begin{eqnarray}
\Gamma[A]&=&-i\ln{\det(i\partial_t-m-A)\over \det (i\partial_t-m)}
\nonumber\\
&=&-i\ln\det\left( 1-S(p)A\right)\nonumber\\
&=&-i{\rm Tr}\ln\left( 1-S(p)A\right)\nonumber\\
&=&i\int_0^\infty {d\tau\over\tau}\; e^{-\tau}\; {\rm Tr}\; e^{\tau
S(p)A}\;+\;{\rm constant}\; .
\end{eqnarray}
Here, the effective action is normalized so that it vanishes when $A=0$. 
Furthermore, $S(p)$ stands for the fermion propagator which, at zero 
temperature, has the form
\begin{equation}\label{propagator}
S(p)={1\over p-m+i\epsilon}
\end{equation}

In the present case, we can identify the proper time Hamiltonian as
\begin{equation}
H=-S(p)A
\end{equation}
and it is straightforward to check that the Heisenberg equations cannot be 
solved in a manner suitable to evaluate the matrix elements. However, let us 
note that equation (\ref{equacao}), in the present case, takes the form
\begin{equation}\label{SA}
\partial_\tau\langle t,\tau\vert t',0\rangle=-\langle t,\tau\vert 
H\vert t',0\rangle=\langle t,\tau\vert S(p)A\vert t',0\rangle\; .
\end{equation}
>From the form of the propagator, at zero temperature, given in 
(\ref{propagator}), we obtain
\begin{equation}
\langle t\vert S(p)\vert t'\rangle=-i\theta(t-t')\; e^{-im(t-t')}\; .
\end{equation}
Therefore, we can also write eq. (\ref{SA}) as 
\begin{eqnarray}
\partial_\tau\langle t,\tau\vert t',0\rangle&=&\int dt''\;
\langle t,\tau
\vert t'',0\rangle\langle t'',0\vert S(p)A\vert t',0\rangle
\nonumber\\
&=&-i\int dt''\;\langle t,\tau\vert t'',0\rangle
\theta(t''-t')A(t')e^{-im(t''-t')}
\; .
\end{eqnarray}
This equation can be integrated to give
\begin{equation}\label{integral}
\langle t,\tau\vert t',0\rangle=\delta(t-t')-
i\int_0^\tau d\tau'\int dt''
\langle t,\tau'\vert t'',0\rangle
\theta(t''-t')A(t')e^{-im(t''-t')}\; .
\end{equation}

In deriving the previous equation, we have used the fact that (see eq. 
(\ref{delta}))
\begin{equation}
\langle t,0\vert t',0\rangle =\delta(t-t')\nonumber
\end{equation}
The equation for the inner product (propagator) is now written as an integral 
equation which cannot always be solved in a closed form. However, we can solve 
the equation iteratively leading to
\begin{eqnarray}
\langle t,\tau\vert t',0\rangle&=&\delta(t-t')-
i\tau\theta(t-t')A(t') e^{-im(t-t')}\nonumber\\
&-&{\tau^2\over 2}\int dt''\;\theta(t-t'')\theta(t''-t')A(t'')A(t')
e^{-im(t-t')}+\;\;...
\end{eqnarray}
We note that the inner product (propagator) in the last equation contains an 
infinite number of terms. However, for equal times, $t=t'$ (which is what we 
need for the trace), all the terms quadratic and higher in the field variables 
vanish because  of opposing theta functions. Consequently, we have
\begin{eqnarray}
\Gamma[A]&=& i\int_0^\infty {d\tau\over\tau} e^{-\tau}\int dt\;\langle t,\tau
\vert t,0\rangle + \mbox{constant}\nonumber\\
&=&{1\over2}\int dt\; A(t)\; .
\end{eqnarray}

This is indeed the exact effective action for this model
\cite{Dunne,DasDunne,DasBar}. Here, we have derived
it, for a first order operator, directly within the proper time formalism by 
solving the integral equation (\ref{integral}). This is quite useful because 
for systems without a closed form expression for the effective action, this 
provides a perturbative expansion of the one loop action. Without going into 
details, we simply note here that at finite temperature \cite{Dasbook}
\begin{equation}
S(p)={1\over p-m+i\epsilon}+2i\pi n_F(m)\delta(p-m)\; ,
\end{equation}
giving
\begin{equation}
\langle t\vert S(p)\vert t'\rangle=-i\left(\theta(t-t')-n_F(m)\right) e^{
-im(t-t')}\; .
\end{equation}
Using this in the equation (\ref{SA}) and various identities derived in 
\cite{DasDunne,DasBar}, one can also obtain the effective action at finite 
remperature in a closed form. But more important is the fact that, even when 
the effective action does not have a closed form, the method gives a 
perturbative expansion of the effective action.

\section{Schwinger Model}

As a final example, we solve the Schwinger Model as well as the general Abelian
 model within the proper time formalism using the method outlined in the 
earlier section. The Lagrangian for the fermions (in $1+1$ dimensional 
Minkowski space) has the form \cite{Schwin62}
\begin{equation}
L_f=\overline\psi\gamma^\mu(i\partial_\mu-eA_\mu)\psi\; .
\end{equation}
Therefore, we obtain the effective action to be 
\begin{eqnarray}\label{Gamma}
\Gamma[A]&=&-i\ln{\det\left( i\partial\!\!\!\slash-eA\!\!\!\!\slash\right)
\over \det(i\partial\!\!\!\slash)}\nonumber\\
&=&-i\ln\det\left( 1-eS(p)A\!\!\!\!\slash\right)\nonumber\\
&=&-i{\rm Tr}\ln\left( 1-eS(p)A\!\!\!\!\slash\right)\nonumber\\
&=& i\int_0^\infty{d\tau\over\tau}e^{-\tau}\; {\rm Tr}\; e^{\tau eS(p)A\!\!\!\!
\slash}\;\; .
\end{eqnarray}
In the present case, 
\begin{equation}
S(p)={p\!\!\!\slash\over p^2+i\epsilon}
\end{equation}
and the \lq\lq Tr"involves a trace over the Dirac indices as well.

Equation (\ref{equacao}), in the present case, takes the form
\begin{equation}
\partial_\tau\langle x,\tau\vert x',0\rangle=-\langle x,\tau\vert H\vert x',0
\rangle\; ,
\end{equation}
with
\begin{equation}
H=-eS(p)A\!\!\!\!\slash\; .
\end{equation}
Following the discussion of the last section, we can write
\begin{eqnarray}
\partial_\tau\langle x,\tau\vert x',0\rangle&=& e\langle x,\tau\vert\; S(p) 
A\!\!\!\!\slash\;\vert x',0\rangle\nonumber\\
&=& e\int d^2x''{d^2p\over (2\pi)^2}\;\langle x,\tau\vert x'',0\rangle\; 
{p\!\!\!\slash A\!\!\!\!\slash(x')\over p^2}e^{-ip\cdot (x''-x')}\;\; .
\end{eqnarray}
Integrating this, as in the last section, we obtain the integral equation
\begin{equation}
 \langle x,\tau\vert x'',0\rangle=\delta^2(x-x')+e\int_0^\tau d\tau'd^2x''
{d^2p\over (2\pi)^2}\langle x,\tau'\vert x'',0\rangle 
{p\!\!\!\slash A\!\!\!\!\slash(x')\over p^2}e^{-ip\cdot (x''-x')}\;\;
\end{equation}
which can be solved iteratively to give
\begin{eqnarray}
\langle x,\tau\vert x'',0\rangle&=&\delta^2(x-x')+\tau e\int 
{d^2p\over (2\pi)^2}
{p\!\!\!\slash A\!\!\!\!\slash(x')\over p
^2}e^{-ip\cdot (x-x')}\nonumber\\
&+&{\tau^2\over2}e^2\int d^2 x''{d^2p\over (2\pi)^2}
{d^2p'\over (2\pi)^2}
{
{p\!\!\!\slash}'A\!\!\!\!\slash(x'')p\!\!\!\slash 
A\!\!\!\!\slash(x')\over {p'}^2 p^2}\; 
e^{-ip'\cdot (x-x'')}e^{-ip\cdot (x''-x')}+\;\; ...
\end{eqnarray}

Once again, we see that the proper time propagator involves an infinite number
 of terms. However, it can be shown easily, following from various
identities 
in $1+1$ dimensions (see \cite{Karev}), that all the higher order terms 
starting with the cubic do not contribute to the trace. In fact, even the 
linear term does not contribute to the trace because of the odd nature of the 
momentum integrand. Thus, we have (\lq\lq tr'' denotes the trace over Dirac 
indices)
\begin{eqnarray}
{\rm Tr}\langle x,\tau\vert x,0\rangle&=& 
{\tau^2\over2}e^2\int d^2 xd^2 x''
{d^2k\over (2\pi)^2}{d^2p\over (2\pi)^2}
tr(\gamma^\mu\gamma^\nu\gamma^\lambda\gamma^\rho)
{(p+k)_\mu p_\lambda\over (p+k)^2 p^2}A_\nu(x'')A_\rho(x)
e^{-ik\cdot(x-x'')}\nonumber\\
&=&-{i\tau^2 e^2\over 2\pi}\int d^2 x 
A_\mu\left( \eta^{\mu\nu}-{\partial^
\mu\partial^\nu\over \partial^2}\right) A_\nu\; .
\end{eqnarray}
Putting this back into eq. (\ref{Gamma}), we obtain
\begin{eqnarray}\label{acaoexata}
\Gamma[A]&=&i\int_0^\infty {d\tau\over \tau} e^{-\tau}\; {\rm Tr}\langle x,
\tau\vert x,0\rangle\nonumber\\
&=& {e^2\over 2\pi}\int d^2 x A_\mu\left( \eta^{\mu\nu}-{\partial^\mu\partial^
\nu\over \partial^2}\right) A_\nu\; .
\end{eqnarray}

This is indeed the exact effective action for the Schwinger model in a gauge 
invariant regularization \cite{DasMathur}(As is well known, there is a one
parameter
arbitrariness depending on the 
choice of the regularization, see ref. \cite{DasMathur} and references
therein. Here we have used a gauge invariant 
regularization for simplicity). Let us next comment on the solubility of the 
general Abelian model in $1+1$ dimensions within this framework without going 
into details. \break

The Lagrangian, for the general model, has the form \cite{Bassetto} (with an
arbitrary  \break parameter $r$)

\begin{equation}
L_f=\overline\psi\gamma^\mu\left( i\partial_\mu -e(1+r\gamma_5)A_\mu\right) 
\psi
\end{equation}
and it reduces to all the $1+1$ dimensional soluble models under different 
limits \cite{Bassetto}. Let us note that in $1+1$ dimensions, the Dirac
matrices satisfy
\begin{equation}
\gamma_5\gamma^\mu=\epsilon^{\mu\nu}\gamma_\nu\; ,
\end{equation}
so that if we define
\begin{equation}\label{B}
B^\mu=(\eta_{\mu\nu}+ r\epsilon^{\mu\nu})A_\nu
\end{equation}
we can write
\begin{equation}
L_f=\overline\psi\gamma^\mu(i\partial_\mu-eB_\mu)\psi\; .
\end{equation}

This is, in fact, the Schwinger model and the effective action would be 
identical 
to (\ref{acaoexata}), in terms of the $B_\mu$ field. Substituting relation 
(\ref{B}), then, would give the exact effective action for the general
Abelian 
model with a gauge invariant regularization.

\section{Conclusion}

We have shown how Schwinger's proper time formalism can be applied directly to 
fermionic systems with first order operators. Here one solves an integral 
equation leading (in general) to a perturbative expansion of the effective 
action. This is, of course, quite useful in odd space-time dimensions where 
the determinant of a first order operator cannot naturally be related 
to that of a second order operator. However, it is  useful in 
even dimensions as well for a perturbative expansion of the effective action.

A.D. would like to thank the members of the Theoretical Physics Department of 
UFRJ for hospitality where this work was done. A.D. is supported in part by US 
DOE Grant number DE-FG-02-91ER40685, NSF-INT-9602559 and a Fulbright grant. 
C.F. is partially supported by CNPq (the National Research Council of 
Brazil).

\end{document}